# Low dimensional neutron moderators for enhanced source brightness

Ferenc Mezei[1,2], Luca Zanini[1], Alan Takibayev[1], Konstantin Batkov[1], Esben Klinkby[1,3], Eric Pitcher[1] and Troels Schönfeldt[1,3]

[1]European Spallation Source, ESS AB, PO BOX 176, 22100 Lund, Sweden
[2]Hungarian Academy of Sciences, Wigner RCF, 1525 Budapest, Pf. 49, Hungary
[3]DTU Nutech, Technical University of Denmark, DTU Risø Campus, 4000 Roskilde, Denmark

Corresponding Author: F. Mezei, European Spallation Source ESS AB, PO BOX 176, 22100 Lund, Sweden, e-mail: ferenc.mezei@esss.se, Tel:+46 72 179 2039

*Abstract*

In a recent numerical optimization study we have found that liquid para-hydrogen coupled cold neutron moderators deliver 3 – 5 times higher cold neutron brightness at a spallation neutron source if they take the form of a flat, quasi 2-dimensional disc, in contrast to the conventional more voluminous shapes used by now. In the present paper we describe a simple theoretical explanation of this unexpected behavior, which is based on the large difference in para-hydrogen between the values of the scattering mean free path for thermal neutrons (in the range of 1 cm) and its much larger equivalent for cold neutrons. This model leads to the conclusions that the optimal shape for high brightness para-hydrogen neutron moderators is the quasi 1-dimensional tube and these low dimensional moderators can also deliver much enhanced cold neutron brightness in fission reactor neutron sources, compared to much more voluminous liquid $D_2$ or $H_2$ moderators currently used. Neutronic simulation calculations confirm both of these theoretical conclusions.

*Keywords*: neutron source, research reactor, spallation source, neutron moderator, cold source, neutron scattering, para-hydrogen

1. Introduction

Neutron moderators for producing the slow neutrons adequate for neutron scattering research are commonly envisaged with dimensions comparable to the neutron scattering mean free path inside the moderator material (e.g. liquid $H_2$, $D_2$, $H_2O$, …). This provides for good efficiency of slowing down the fast or partially slowed down neutrons arriving primarily from the reflector surrounding the moderator and achieving a reasonably homogeneous distribution of the moderated neutrons inside the moderator volume. Actually the highest equilibrium moderated neutron flux is expected to occur well inside the volume of the moderator in view of the leakage of slow neutrons at the surface. Typical examples are cold neutron sources at fission reactor facilities with usual reflector materials such as Be or heavy water $D_2O$ and both thermal and cold moderators at spallation sources. A major recent development at spallation sources is the use of a few cm

H$_2$O premoderator around the cold moderators, originally pioneered by Noboru Watanabe and coworkers (see [1] and references therein). The neutron absorption by the moderators is not fully negligible, but its mean free path is considerably longer than the scattering mean free paths in moderator materials, and can be neglected in the fundamental considerations that follow.

A remarkable deviation from the envisaged homogeneity of flux distribution inside the moderator has been observed in the simulation calculations on the coupled cold moderator at the MLF spallation neutron source facility at J-PARC [1]. In liquid para-H$_2$ moderator with H$_2$O water premoderator the flux distribution leaving the moderator has shown a clear maximum near to the walls when projected back onto the moderator, cf. Fig. 16 in [1]. This observation turned out to have far-reaching relevance for the recent developments described below.

2. Model for neutron moderation in para-hydrogen

The above anomaly can be explained by a simple qualitative model based on the fact that in para-H$_2$ the scattering mean free path $L_T$ for thermal neutrons is much shorter than the scattering mean free path $L_C$ for cold neutrons [2]. The thermal neutrons from the water premoderator get slowed down by the first collision close to the moderator wall after entering the para-H$_2$ and they can escape from the moderator volume as cold neutrons without any further collision, since $L_C >> L_T$. Therefore the cold neutron density inside the moderator remains higher within a layer of thickness about $L_T$ near the moderator walls. Emission in a given direction at a given point on the moderator surface will reflect the moderated neutron density integrated over the backward extension of the neutron path from the surface into the moderator volume ("emission path"), taking into account attenuation corresponding to the mean free path $L_C$. If this path goes through the central part of the moderator volume, it will display high brightness portions of length $L_T$ when crossing the moderator walls and lower brightness inside the moderator volume. If the emission path runs parallel to a moderator wall in close range within distance $L_T$, the length of its effective bright portion can be as large as $L_C$, and the emitted intensity will be enhanced of the order of the ratio $L_C/L_T$. The increased brightness revealed in Fig. 16 of [1] at the edges of the moderator can be explained by this mechanism, and its dependence on the distance from the moderator edge provides the estimate $L_T \approx 1$ cm.

3. The concept of low dimensional moderators

Recently we have investigated by MCNPX simulations the neutron brightness of cylinder shaped para-H$_2$ coupled cold neutron moderators at spallation sources under the "unperturbed" condition of being fully surrounded by reflector and / or light water premoderator, i.e. without openings for the extraction of beam lines [3]. The results show that the neutron emission brightness of the cylindrical side of the moderator increases as the height of the cylinder is reduced, i.e. the moderator shape becomes a disc.. The simple model discussed above offers a way to understand the physical mechanism behind these simulation results. The emission paths for neutrons emerging on the edge of the flat disc pass in close distance $< L_T$ parallel to at least one of the two circular faces of the moderator over distances much larger than $L_T$. Therefore for the neutron emission we see a much larger bright moderator depth along the emission path compared to the conventional moderator geometries. This explains the observed enhanced moderator brightness.

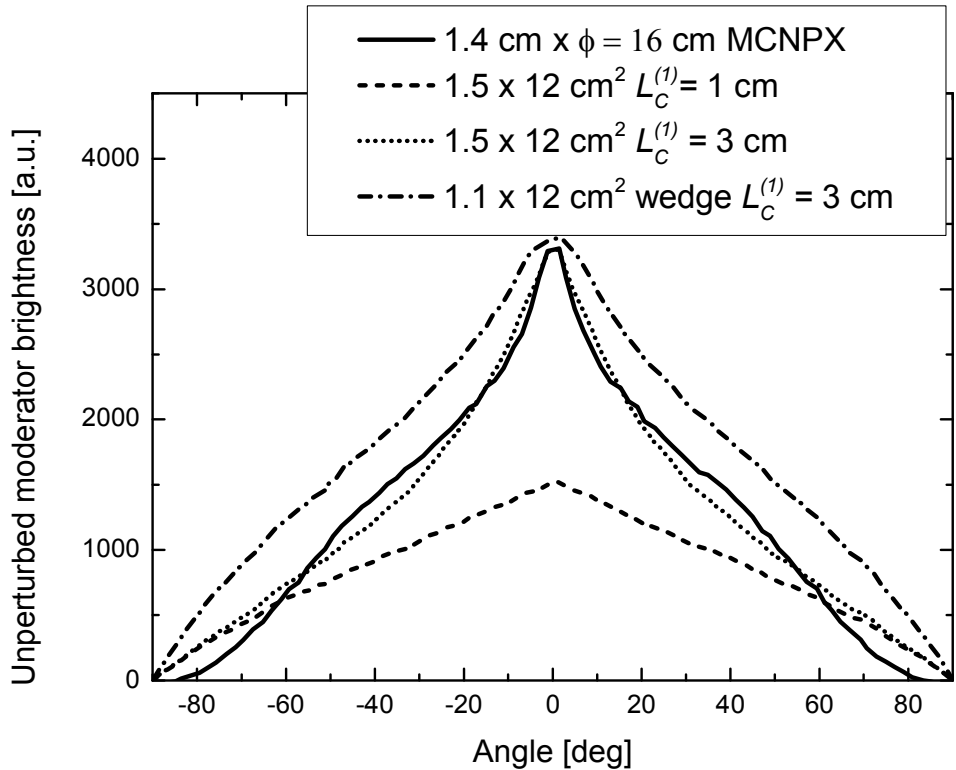

Fig. 1 Calculated angular dependence of the neutron brightness as a function of the neutron emission angle with respect of the plane of flat para-$H_2$ moderators. The solid line shows the MCNPX results for a 1.4 cm high, 16 cm diameter disc shaped moderator [3]. The other distributions have been calculated by the schematic single collision model described in the text for the rectangular neutron emitting surfaces and $L_C^{(1)}$ effective attenuation constants as indicated in the figure. The depth of all model moderators is 16 cm, and details on the wedge shaped one are given in the text.

The above schematic single collision model can remarkably reproduce some quantitative features of the results in [3], assuming homogeneous reflector-premoderator flux distribution, exponential decay of the thermal neutron flux with the distance from the moderator walls (effective decay constant $L_T^{(1)}$) and exponential attenuation of the cold neutron propagation through the moderator material with an effective decay constant $L_C^{(1)}$. Namely, we have observed that the angular distribution of neutron emission in the vertical direction (i.e. that of the moderator cylinder axis) reveals a collimated behavior for the 1.4 cm high flat moderator, with a sharp peak in direction of the center plane of the moderator disc. This distribution could be reproduced in a good approximation by the present model with parameters $L_T^{(1)} = 1$ cm (the value suggested by that data in [1]) and $L_C^{(1)} = 3$ cm, as shown in Figure 1. Actually the shape of the curve is rather sensitive to the choice of $L_C^{(1)}$ as illustrated by the curve calculated for $L_C^{(1)} = 1$ cm. For higher values of $L_C^{(1)}$ the peak brightness is getting steeply higher and the angular distribution narrower around angle 0, corresponding to larger bright moderator depth seen through a finer collimation defined by the moderator walls. The estimate of $L_C^{(1)}/L_T^{(1)} \sim 3$ illustrates well within the single collision model the role of the difference between both mean free paths. Since multiple collisions tend to strongly broaden the angular distribution, the approximate value $L_C^{(1)} = 3$ cm is a lower limit for the true mean free path $L_C$ (known to be $\sim 11$ cm). Making the vertical cross section of the moderator wedge shaped allows us to widen the angular distribution of the emitted beam. This is illustrated by the dot-dashed curve in Fig. 1 for a trapezoidal moderator cross section with 1.1 cm height at the neutron emitting surface and 3.1 cm at the back side at 16 cm distance (compared to the 1.5 cm x 16 cm cross section for the flat disc, dotted curve).

4. Model predictions and confirmation by MCNPX simulations

These observations and considerations suggest that the difference between thermal and cold neutron mean free paths $L_T$ and $L_C$, respectively, plays a decisive role in the 3 -5 times higher brightness of flat coupled liquid para-$H_2$ moderators at a spallation source compared to the conventional design, as discovered in [3]. (There are in addition other factors favourable for the brightness performance of low dimensional moderators: they can better be concentrated to the maximum flux zone in the reflector, their smaller neutron emitting surface necessitates smaller amount of reflector to be removed for openings for neutron beams.)  The difference between these two mean free paths removes constraints analogous to Liouville theorem in neutron propagation. The basic idea of taking advantage of such a difference is to surround long neutron emission paths with walls of the moderator. Thus all the depth corresponding to $L_C$ contributes to the neutron emission, while the slowing down of the thermal neutrons happens over a distance $L_T$. When the two mean free paths are equal (as in ortho-$H_2$), we can only see into an emission depth $L_T$. In the flat disc moderators proposed in [3] is the extended emission path is achieved by the essentially 2-dimensional shape of the moderator, with the two parallel circular walls close to each other. Along the same lines of logics, one can predict that:
    a) Highest brightness is to be expected from quasi 1-dimensional "tube" moderators (instead of quasi 2-dimensional flat moderators), where the emission paths are surrounded by moderator walls on all sides, e.g. by 1.5 cm x 1.5 cm viewed surface at the end of a 15 – 30 cm tube.
    b) Low dimensional para-$H_2$ moderators should provide higher cold neutron brightness at reactor sources compared to the by now used voluminous $D_2$ and $H_2$ moderators, in similar proportions to the gains reported in [3] for coupled cold spallation sources moderators.
    c) The brightness of flat moderators is expected to have its maximum in the limit of 0 thickness. This has been actually corroborated in the meantime by simulation calculations for the perturbed flux (to be published). The 1.4 cm optimum in [3] appears to be artificial, due to tallying within ± 5° emitted beam divergence, while in the 0 thickness limit the emitted beam divergence tends to 0°.

Our latest MCNPX simulations have confirmed both of expectations a) and b). The perturbed source brightness was evaluated for various moderator models placed at 55 cm from the center of an ILL like enriched uranium fission zone of 39 cm diameter and 80 cm height, within a 2 m diameter heavy water reflector tank. Point detector tally was used at 5 m distance in tangential beam lines. As shown in Fig. 2, the "flat" and "tube" type low-dimensional moderators provide close to an order of magnitude gain in brightness around 2.5 Å neutron wavelength, with the highest brightness achieved by the tube moderator. The gain in brightness drops to some 75 % for above 6 Å. In practice an additional gain in performance of > 60 % can be expected, since the about 400 times smaller volume tube moderator can be placed closer to the reactor core by at least 20 cm than the center of the optimized sized large $D_2$ moderator.

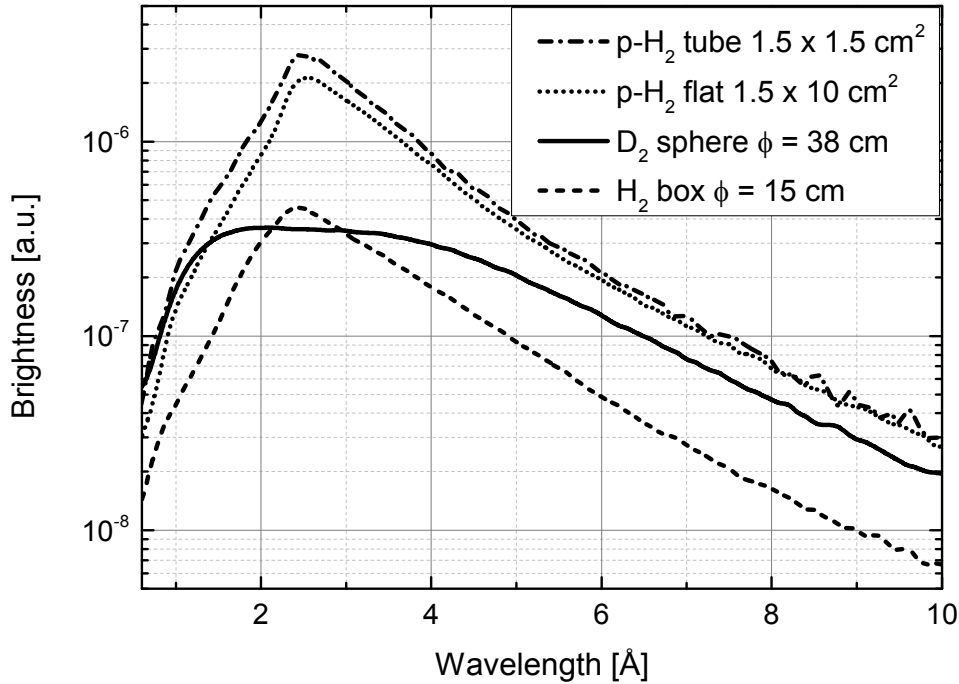

Fig. 2. MCNPX simulation results for the relative perturbed neutron brightness of a variety of cold neutron moderators at an ILL type fission reactor with heavy water reflector. The dimensions in the figure define the viewed moderator surface. The depth of the rectangular flat and tube moderators is 25 cm. The $H_2$ moderator contains 25 % ortho and 75 % para, and it is 5 cm thick. More details on the geometry are given in the text.

As shown in [3] a flat disc moderator offers a large angular distribution of emitted neutron brightness in the plane of the disc and is adequate to serve beam lines distributed over a large angular range in this plane. The angular distribution for a tube moderator has a strong peak along the axis (similar to the dotted curve in Fig. 1), which can be broadened somewhat by a wedge/conical shape tube as described above in connection with the figure. One end of a tube moderator can serve a bundle of neutron guides within a 15 – 20° angular range, and one will need an array of tube moderators for a set of beam lines distributed over a large angular range. Such a set could be combined into a star shaped moderator with some 10 points. Remarkably, with para-$H_2$ (and eventual other similar cases of dispersive neutron propagation) optimal moderator brightness is achieved by shapes aimed at enhancing the surface to volume ratio for better access of volumes near moderator walls, in contrast to the re-entrant structures used for other moderators aimed at accessing the center of the volume.

5. Conclusions

We can conclude that for a material that displays a larger mean free path for the moderated neutrons than for the more energetic neutrons delivered by the surrounding reflector / premoderator system low dimensional moderator geometries offer higher efficiency in terms of moderator brightness for all neutron sources than the now common moderator design. The smaller volume of low dimensional moderators also reduces the power of cryogenic cooling required, and can allow for additional gains in performance by placing the small volume moderators closer to the source of fast neutrons. By this the total gain in brightness by tube moderators at an ILL type reactor is expected to exceed the factor of 10 around 2.5 Å wavelength and a factor of 2.5 above 6 Å. This is a most significant new opportunity in neutron research and facility design.

In particular, in view of Liouville theorem, it is the moderator brightness that limits the neutron flux that can at best be delivered to the samples. It is straightforward to demonstrate that the Liouville limit can be achieved with reasonable efficiency by established neutron guide design for sample areas some 30 – 50 % smaller than the viewable moderator surface. The gain in brightness by a substantial factor might also compensate for the less efficient brightness transmission to sample areas moderately larger than the emitting moderator surface. The definition of the optimal moderator lay-out for a facility with a large variety of neutron scattering instruments and sample sizes is a complex design task, which is beyond the scope of this paper.

Acknowledgements

The authors are grateful for inspiring discussions with John Haines, Phil Ferguson and Franz Gallmeier and many other colleagues in the neutron scattering community.